\documentclass[useAMS,usenatbib]{mn2e}
\usepackage{natbib}

\usepackage{graphicx}
\usepackage{epsfig}

\usepackage{txfonts}

\usepackage{hyperref}

\usepackage{float}
\usepackage{color}
\usepackage{lscape,graphicx}
\usepackage{journal_shortcuts}
\usepackage{amssymb}
\usepackage{multirow}
\usepackage{afterpage}
\usepackage{ulem}
\usepackage{threeparttable}
\title[NIR Virial Black Hole mass estimates]{
Extending Virial Black Hole Mass Estimates to Low-Luminosity or Obscured AGN: the cases of NGC 4395 and MCG -01-24-012}
\author[F. La Franca et al. ]{F.~La~Franca,$^{1}$\thanks{E-mail:
lafranca,onori,riccif@fis.uniroma3.it} 
F.~Onori,$^{1}$ 
F.~Ricci,$^{1}$
E.~Sani,$^{2}$
M.~Brusa,$^{3}$$^,$$^{4}$ R.~Maiolino,$^{5}$
S.~Bianchi,$^{1}$ 
  \newauthor A.~Bongiorno,$^{6}$ F.~Fiore,$^{6}$ A.~Marconi$^{7}$ 
and C.~Vignali$^{3}$$^,$$^{4}$\\
$^{1}$Dipartimento di Matematica e Fisica, Universit\`a Roma Tre,
   via della Vasca Navale 84, 00146 Roma, Italy\\
$^{2}$INAF - Osservatorio Astrofisico di Arcetri, largo E. Fermi 2, 50125 Firenze, Italy\\
$^{3}$Dipartimento di Fisica e Astronomia, Universit\`a di Bologna, viale Berti Pichat 6/2, 40127 Bologna, Italy\\
$^{4}$INAF - Osservatorio Astronomico di Bologna, via Ranzani 1, 40127 Bologna, Italy\\
$^{5}$Cavendish Laboratory, University of Cambridge, 19 J. J. Thomson Ave., Cambridge CB3 0HE, UK\\
$^{6}$INAF - Osservatorio Astronomico di Roma, via Frascati 33, 00044 Monte Porzio Catone, Italy\\
$^{7}$Dipartimento di Fisica e Astronomia, Universit\`a di Firenze, largo E. Fermi 2, 50125 Firenze, Italy}
\begin{document}

\date{Accepted 2015 February 19. Received 2015 February 17; in original form 2014 July 11}

\pagerange{\pageref{firstpage}--\pageref{lastpage}} \pubyear{2014}

\maketitle

\label{firstpage}

\begin{abstract}
In the last decade, using single epoch (SE) virial based spectroscopic optical observations, it has been possible to measure the  black hole (BH) mass on large type 1 Active Galactic Nuclei (AGN) samples. However this kind of measurements can not be applied on those obscured type 2 and/or low luminosity AGN where the nuclear component does not dominate in the optical. 
We have derived  new SE relationships, based on the FWHM and luminosity of the broad line region component of the Pa$\beta$ emission line and/or the hard X-ray luminosity in the 14-195 keV band, which have the prospect of better working with low luminosity or obscured AGN.
The SE relationships have been calibrated in the $10^5 - 10^9$ ${\rm M}_{\odot}$ mass range, using a sample of  AGN whose BH masses have been previously measured using reverberation mapping techniques. 
Our tightest
relationship between the reverberation-based BH mass and the SE virial
product has an intrinsic spread of 0.20 dex.

Thanks to these SE relations, in agreement with previous estimates, we have measured a  BH mass of M$_{\rm BH}$=$1.7^{+1.3}_{-0.7} \times 10^5$  M$_{\odot}$ for the low luminosity, type 1, AGN NGC 4395 (one of the smallest  active galactic BH known). We also measured, for the first time, a BH mass of M$_{\rm BH}$=$1.5^{+1.1}_{-0.6} \times 10^7$  M$_{\odot}$ for the Seyfert 2 galaxy  MCG -01-24-012.

\end{abstract}


\section{Introduction}

Nowadays there is robust evidence that almost every galaxy host a super
massive black hole (BH; M$_{\rm BH}$=10$^{5}$-10$^{9}$ M$_{\odot}$) and
that many
correlations exist between the BH mass and some of the host galaxy properties 
\citep[e.g. bulge mass, luminosity and stellar dispersion;][]{ferrarese00, gebhardt00,marconihunt03,sani11}.
Then, it has been argued that the
existence of these scaling relationships implies that the evolution of
galaxies and the growth of the super massive BHs (SMBH) are tied together (the AGN/galaxy co-evolution scenario).

In the last decade, using virial based techniques in the optical band, it has been possible to measure the BH mass on large type 1 AGN (AGN1) samples and therefore derive the super massive BH mass function \citep{greene07b,kelly09,kelly10,merloni10,bongiorno14}.  Many of these measures are based on the Single Epoch (SE) BH mass estimates.
By combining the velocity of the Broad Line Region (BLR) clouds (assuming Keplerian orbits) along with their distance R it is possible to determine the total mass contained within the BLR (which is dominated by the BH) in a simple way using
\begin{equation}
{\rm M}_{\rm BH}= {{f \Delta {\rm V}^2 {\rm R}}\over{\rm G}},
\end{equation}
where G is the gravitational constant and {\it f} is a factor that depends on the geometric and kinematic structure of the BLR (e.g., Kaspi et al. 2000).
These techniques derive the AGN BH mass using SE spectra to measure $\Delta$V from the FWHM of 
some of the BLR lines (typically: H$\beta$ or MgII $\lambda$2798 or CIV $\lambda$1459) and R from either the continuum or the line luminosities, L, which have been proved to be proportional to R$^2$ \citep[i.e. $R\propto L^{0.5}$;][]{bentz06b}. Therefore, the SE estimates are based on relations of the type

\begin{equation}
{\rm log M}_{\rm BH}  =  {\rm log (  FWHM^\alpha \cdot L^\beta )} + \gamma, 
\label{Mrev}
\end{equation}

\noindent
where 
$\alpha\sim 2$, $\beta\sim 0.5$ \citep[e.g.][but see also \citealt{dietrich04}]{wandel99, vestergaard02, vestergaard04, vestergaard09, mclure02}. These relationships have typical spreads of $\sim$0.5 dex.

However these measurements can not be applied on those obscured or partially obscured sources where the AGN component
does not dominate in the optical: type 2 AGN (AGN2), intermediate class AGN (AGN1.8 and AGN1.9) or even AGN1 having low luminosities and low mass  BH (M$_{\rm BH}$$\le$10$^{6}$ M$_{\odot}$). Indeed, in these cases, in the (rest frame) optical band,  either the  broad line component is not visible or, in any case, the AGN continuum and line luminosities are affected by severe obscuration and galaxy contamination thus preventing  a reliable estimate of the BH mass.
 
There are, in summary, indications that the SE techniques do not allow
to obtain an un-biased estimate of the
density of the low-luminosity (and, therefore, low-mass; M$_{\rm BH}$$\le$10$^{7}$ M$_{\odot}$) class of AGN. 
Such a bias could have strong implications
on our understanding on the AGN/galaxy evolution. 

Indeed, the density of
low mass BHs  in the local universe provide
unique tests for studies of BH formation and growth, galaxy
formation and evolution. In current models of galaxy evolution in a
hierarchical cosmology, SMBHs must have been built up from accretion
onto much smaller seeds, in conjuction with merging with other
BHs. These models also predict that smaller scale structures form at
later times (cosmic downsizing), and one might expect that seeds BHs
in these smaller systems may not have had enough time to be fully
grown. This means that low-mass BHs likely contain clues about the
formation of the first black holes, therefore the mass function of the
present day low-mass black holes and their host galaxies properties
can be used to discriminate between different models for seeds BHs and
help shed light on the coevolution of BHs and galaxies 
\citep[see][]{dong06,greene07a,dong12,greene12}.

In the last few years \citet{landt08,landt11b,landt13} have studied quasi-simultaneous near-IR (NIR) and optical spectra of
a sample of well-known broad-emission line AGN, whose BH masses were already determined. 
 These studies have allowed the calibration of a NIR relationship for estimating AGN BH masses based on the 
 widths of the Paschen hydrogen broad emission lines and the total 1 ${\rm \mu}$m continuum luminosity \citep{landt11b,landt13}.
 However the virial relationship developed in \citet{landt11b,landt13} makes use of the 1 ${\rm \mu}$m continuum luminosity, which 
 has to be derived by a Spectral Energy Distribution (SED) fitting method and can suffer from galaxy contamination especially for intrinsic low-luminosity AGNs. 

Following previous studies in which either a correlation between the BLR radius and the absorption corrected X-ray luminosity was found \citep[][see also \citet{maiolino07}]{greene10b}, or
the luminosity of the BLR emission lines can be used as a proxy of the BLR radius \citep{greene05},
we have  decided to calibrate new virial relationships based on the ${\rm Pa\beta}$ broad emission line FWHM combined
with either the hard X-ray or the broad emission line luminosities \citep[see also ][for similar SE methods based on the Pa$\alpha$ and Pa$\beta$
emission-line luminosities]{kim10}. Such relations are potentially able to be applied 
on both the low-luminosity and the optically more obscured AGN. The method has been successfully tested on one of the lowest BH mass Seyfert 1 known, NGC 4395, and on a Seyfert 2 galaxy, MCG -01-24-012; both
spectroscopically observed in the NIR in the framework of a large programme  aimed at measuring
the BH mass of low-luminosity or type 2 AGN (Onori et al., in preparation), included in the  22 month Swift/Burst Alert Telescope (BAT) catalogue \citep{tueller10}.

 Unless otherwise stated, all quoted errors are at the 68\% (1 $\sigma$) confidence level. 
 We assume ${\rm H_{0}=70}$ ${\rm km \, s^{-1} Mpc^{-1}}$, ${\rm \Omega_{m}=0.3}$ and ${\rm \Omega_{\Lambda}=0.7}$.
 
 \section{NGC 4395 and MCG -01-24-012 and their previous BH mass measures}\label{sec:NGC4395}
 
 \subsection{NGC 4395}
 
NGC 4395 ($\alpha$= 12$^h$ 52$^m$ 48.8$^s$, $\delta$ = 33$^\circ$ 32$^m$ 49$^s$; J2000) is a small bulgeless (Sd) galaxy hosting one of the smallest, low luminosity, active super massive BH ($10^4-10^5$ ${\rm M_{\odot}}$) ever found\footnote{One other
 small active BH, of similar size, is that of POX 52: an AGN1 in a dwarf elliptical galaxy, whose BH mass, according
 to the extrapolation of the M$_{BH}$-$\sigma$ relation \citep{tremaine02} and to the SE, H$\beta$ based, relationship
 is expected to have ${\rm M_{BH} \sim 1.3-1.6  \times 10^5 {\rm M}_{\odot}}$ \citep{kunth87, barth04, barth05}.
}.  Its 14-195 keV X-ray luminosity is ${\rm L_X = 10^{40.79}  ~erg~ s^{-1}} $ \citep{baumgartner13}.
The optical spectrum (Figure \ref{fig:OptSpct}) reveals very strong narrow emission lines showing weak broad wings in the H$\alpha$ and H$\beta$ emission lines. Based on the relative intensities of the narrow and broad components, \citet{filippenko89} originally classified the nucleus of NGC 4395 as a type 1.8 or type 1.9 Seyfert. Later, using HST spectroscopy, the nucleus of NGC 4395 has been classified as a type 1 Seyfert \citep{filippenko93}. According to \citet{ho97IV}, the BLR component of the H$\alpha$ line has a FWHM=442 km s$^{-1}$.

Different methods have been employed in the last years to derive the central BH mass of NGC 4395.
 From photoionization modeling
of the narrow-line region and BLR, \citet{kraemer99}
found  that the broad H$\beta$ line originates from a region
with radius $R =3\times10^{-4}$ pc. Assuming that the line
emitting gas is gravitationally bound, and using the measure of the width 
of the H$\beta$ BLR component, FWHM = 1500 km s$^{-1}$ \citep{kraemer99}, they found
$\rm M_{BH}\simeq 1.2 \times 10^5$ M$_\odot$. A result similar to \citet{peterson05}, who obtained 
${\rm M_{BH} = (3.6 \pm 1.1) \times 10^5}$ M$_{\odot}$, 
through reverberation mapping of the CIV line. This last estimate has been halved to $(1.8 \pm 0.6) \times 10^5$ M$_{\odot}$ due to the updated 
 virial {\it f}-factor ($f = 4.31$) in \citet{graham11}. The last up to date broadband photometric reverberation, made by \citet{edri12} 
 using the H$\beta$ line,   has found an even lower mass of  ${\rm M_{BH} = (4.9\pm2.6) \times 10^4}$ M$_{\odot}$;  
 this is essentially only due to the much smaller geometrical $f$-factor used ({\it f} = 0.75 instead of {\it f} = 5.5 used by \citealt{peterson05}). Indeed,
 if a common geometrical factor $f=4.31$ is used (see next section), all the above measures agree to a BH mass of ${\rm M_{BH} \sim 2.7 \times 10^5}$
 M$_{\odot}$.
  The lack in this galaxy of a significant bulge and the stringent upper limit of ${\rm 30\, km\,s^{-1}}$ on its velocity 
 dispersion confirm a value of the order of $10^4-10^5$ M$_{\odot}$ on its BH mass \citep{filippenkoho03}.
 
As the H$\alpha$ line shows a FWHM=442 km s$^{-1}$ and a  luminosity
 of 10$^{38.08}$ erg s$^{-1}$ \citep{ho97IV}, in principle, a SE relationship, based on the two above quantities could be used to derive the  BH mass. We used eq. A1 in \citet{greene07b}, but converted to a geometrical virial factor $f=4.31$. 
 In this case it results a mass of $5.5\times10^4$ M$_\odot$, which is smaller than previously measured.  This lower value could be caused by extinction and dilution effects from the hosting galaxy, which are more relevant in small mass and  low luminosity AGNs, such as NGC 4395. Nonetheless, the observed extinction
 is not able to correct the H$\alpha$ luminosity in order to obtain a BH mass in agreement with the previous described measures. In fact, from the He II $\lambda$1640/$\lambda$4686 ratio a reddening of $E_{B-V}=0.05$ mag is found \citep{kraemer99}, while from 
 to the observed Balmer decrement of the narrow line component \citep[see Table 1 in ][]{kraemer99}, a reddening  $E_{B-V}=0.08$ mag  is measured. Assuming a Galactic extinction law (R$_V$=3.1), these values correspond to an extinction of $\sim$0.09 and $\sim$0.14 mag, respectively, at the H$\alpha$ wavelength. 
In this case, once the H$\alpha$ luminosity has been corrected for this small extinction, a BH mass of $\sim 5.7\times10^4$ M$_\odot$ is obtained. This low value could be due to an under-estimate of the extinction, which in the inner regions of the galaxy, could be larger than measured from the narrow line region. These results support our project of deriving  new BH mass virial relationships based on NIR and/or hard X-ray luminosity measures which have the prospect of being less affected by extinction and host galaxy contaminations.

 \subsection{MCG -01-24-012}

MCG -01-24-012  ($\alpha$= 9$^h$ 20$^m$ 46.2$^s$, $\delta$ = -8$^\circ$ 3$^m$ 22$^s$; J2000) is a nearby spiral galaxy, at redshift z=0.0196, which hosts a Seyfert 2 nucleus \citep{degrijp92}. 
Using BeppoSAX/PDS telescope  observations it was identified by \citet{malizia02} as the counterpart of the X-ray source  H0917-074, detected by HEAO1/A2 \citep{piccinotti82}. 
The 2-10 keV flux, measured by BeppoSAX, is $\sim1\times10^{-11}$ erg s$^{-1}$ cm$^{-2}$, while the X-ray spectrum shows the presence of iron 
K$\alpha$ emission lines, together with an absorption feature at $\sim8.7$ keV which cannot be explained by the presence of warm material around the source. It turned out to be Compton-thin, having N$_{\rm H}\sim7\times10^{22}$ cm$^{-2}$. Its 14-195 keV luminosity is ${\rm L_X = 10^{43.55}~erg~ s^{-1}}$
\citep{baumgartner13}
and the [O III] image shows an  extended (1.15$''$$\times$2.3$''$; 460 pc $\times$ 910 pc) emission, with the major axis along PA=75 deg, interpreted as an extended NLR \citep{schmitt03}. MCG -01-24-12 has also been observed by Spitzer/IRS. The corresponding low resolution staring-mode spectrum exhibits deep silicate absorption at 10 $\mu$m and weak PAH emission \citep{mullaney11}. 

\begin{figure}
   \centering
   
   \includegraphics[width=0.99\hsize]{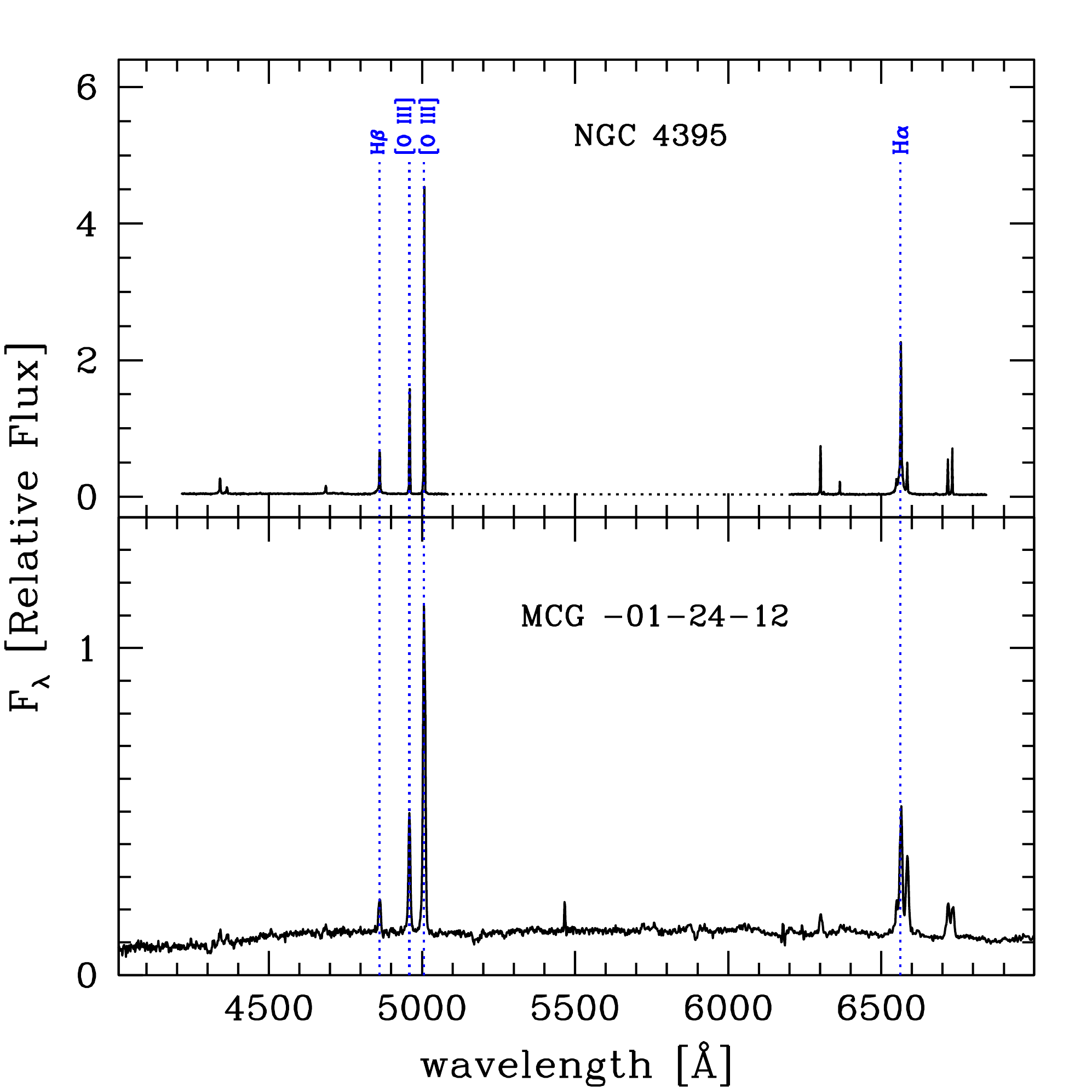}
      \caption{{\it Top.} Red and blue, rest frame, optical spectra of NGC 4395 taken with the double spectrograph on the Hale 5 m telescope \citep{ho95}.
      {\it Bottom.} Rest frame optical spectrum of MCG -01-24-12 taken by the 6dF Galaxy Survey \citep{jones09}. The positions of the H$\alpha$, H$\beta$
 and [OIII] lines are shown.             }
         \label{fig:OptSpct}
   \end{figure}

Using K-band stellar luminosity of the bulge, Winter et al. (2009) derived a black hole mass log(M$_{\rm BH}$/M$_\odot$) = 7.16. The optical spectrum (Figure \ref{fig:OptSpct}), obtained by the 6dF Galaxy Survey \citep{jones09}, shows narrow H$\beta$ an H$\alpha$ emission lines (but see also discussion in sect. \ref{SectMassMCG} about a possible presence of a faint broad H$\alpha$ component).

 \section{Single Epoch Mass relations in the Near-Infrared}
 
 \subsection{Calibration sample}  
 
 As previously discussed,
 we are interested in deriving new BH mass virial relationships based on the
 measure of the width of the BLR component in the Pa$\beta$ emission line. In order not to suffer from problems related to 
 the galaxy contamination, we have tried to substitute the 1 $\mu$m continuum luminosity used by \citet{landt13}
 as the BLR size indicator, with either the hard X-ray 14-195 keV band luminosity, ${\rm L_{X}}$, such as  measured by the 70 months Swift/BAT 
 survey by integrating within the years 2005 and 2010 \citep{baumgartner13}, or the luminosity of the Pa$\beta$, L$_{\rm Pa\beta}$,
 itself (see \citet{kim10}, for a similar study based on the Pa$\alpha$ and Pa$\beta$ emission-line luminosities).

 The calibration sample contains those 20 AGN1 whose broad Pa$\beta$ line fits have been published by  \citet{landt08, landt13} and whose
 BH mass have been measured by the reverberation mapping techniques. When available, we have used 
 the BH mass values listed in \citet{grier13} where the new estimate of the geometrical $f$-factor $f = 4.31$ was adopted.
All the remaining BH masses, which were originally published using $f=5.5$,  have then been converted
 to $f=4.31$. All but one the sources  are listed in the 70 months Swift/BAT catalogue \citep{baumgartner13} from which the 14-195 keV hard X-ray luminosities have been taken.
 In Table \ref{table:1} we present the main characteristics of our calibration sample, such as: redshift, 
 optical spectroscopic classification, BH mass and the relative reference, 14-195 keV hard X-ray luminosity and 
 the FWHM of the broad component of the Pa$\beta$. 
 The FWHMs have been corrected for instrumental resolution using the values reported in \citet{landt08,landt13}. 
 In particular the source 3C 120 has been observed with an average spectral resolution of FWHM ${\rm \sim 180}$ ${\rm km\,s^{-1}}$, 
 while Mrk 279, NGC 3516, NGC 4051 and PG 0844+349 spectra have a resolution of FWHM ${\rm \sim 400}$ ${\rm km\,s^{-1}}$ \citep{landt13}. 
 The instrumental resolution for the remainder of the sources is 
 FWHM ${\rm \sim 350}$ ${\rm km\,s^{-1}}$ \citep{landt08}. 
%
\begin{table*} 
 \centering
 \begin{minipage}{140mm}
\caption{
Calibration Sample.}
\label{table:1}
\begin{center}
\begin{tabular}{@{}lclccccc}
\hline
Object Name & z & class &${\rm M_{BH}}$ & ref & 
$\log{\rm L_{\rm X}}$ & ${\rm FWHM_{Pa\beta}}$ & $\log{\rm L_{\rm Pa\beta}}$\\
 & & & [${\rm M_{\odot}}$] & & [${\rm erg \, s^{-1}}$] & [${\rm km\,s^{-1}}$] &  [${\rm erg \, s^{-1}}$]\\
(1) & (2) & (3) & (4) & (5) & (6) & (7) & (8)\\
\hline
3C 120  & 0.0330 & S1.0 & 5.26$\pm$0.52\, $ \cdot 10^7$            & G13 & 44.38$\pm$0.01 & 2727 &...\\
3C 273  & 0.1583 & S1.0 & 6.94$\pm$1.47\, $ \cdot 10^8$            & P04 & 46.48$\pm$0.03 & 2895 & 43.64\\
Ark 120 & 0.0323 & S1.0 & $101^{+17}_{-25}$~~~\,~~\,\, $\cdot 10^6$       & G13 & 44.23$\pm$0.02 & 5102 & 42.12\\
Mrk 79  & 0.0222 & S1.2 & $8.28^{+1.9}_{-3.2}$\,\, ~~~$\cdot 10^7$         & G13 & 43.72$\pm$0.02 & 3506 & 41.28\\
Mrk 110 & 0.0353 & S1n  & $2.24^{+0.56}_{-0.91}$\,\, ~~$\cdot 10^7$        & G13 & 44.22$\pm$0.02 & 1886 & 41.60\\
Mrk 279 & 0.0304 & S1.0 & 3.10$\pm$0.47\, $ \cdot 10^7$          & G13 & 43.92$\pm$0.02 & 3546 &...\\
Mrk 290 & 0.0296 & S1.5 & 1.90$\pm$0.29\, $ \cdot 10^7$          & D10 & 43.67$\pm$0.04 & 4228 & 41.46\\
Mrk 335 & 0.0258 & S1n  & 1.11$\pm$0.29\, $ \cdot 10^7$          & P04 & 43.45$\pm$0.05 & 1825 & 41.37\\
Mrk 509 & 0.0344 & S1.5 & 9.57$\pm$0.43\, $ \cdot 10^7$          & G13 & 44.42$\pm$0.01 & 3057 & 42.12\\ 
Mrk 590 & 0.0264 & S1.0 & $3.15^{+0.52}_{-0.69}$\,\, ~~$\cdot 10^7$          & G13 & 43.42$\pm$0.07 & 3949 & 40.33\\
Mrk 817 & 0.0314 & S1.5 & $6.29^{+0.95}_{-1.08}$ \,\,~ $\cdot 10^7$ & G13 & 43.77$\pm$0.03 & 5519 & 41.47\\ 
Mrk 876 & 0.1290 & S1.0 & 2.19$\pm$1.01\, $\cdot 10^8$          & P04 & 44.73$\pm$0.07 & 6010 & 42.50\\
NGC 3227& 0.0039 & S1.5 & $2.24^{+0.86}_{-0.91} $\,\,~~ $\cdot 10^7$ & G13 & 42.56$\pm$0.01& 2934 & 40.03\\
NGC 3516& 0.0088 & S1.5 & $3.10^{+0.30}_{-0.26}$ \,\,~ $\cdot 10^7$ & G13 & 43.31$\pm$0.01 & 4451 &...\\
NGC 4051& 0.0023 & S1.0 & $2.2^{+0.2}_{-0.4}$\, ~~~~~~$\cdot 10^6$ & G13 & 41.67$\pm$0.02 & 1633 &...\\
NGC 4151& 0.0033 & S1.5 & $3.62^{+0.39}_{-0.22}$ \,\,~ $\cdot 10^7$ & G13 & 43.12$\pm$0.01 & 4654 & 40.43\\
NGC 4593& 0.0090 & S1.0 & $9.1^{+1.7}_{-1.3}$\,  ~~~~~~$\cdot 10^6$            & G13 & 43.20$\pm$0.01 & 3775 & 40.68\\
NGC 5548& 0.0172 & S1.5 & 5.95$^{+0.73}_{-0.86}$\, \,~~$\cdot 10^7$ & G13 & 43.72$\pm$0.01 & 6516 & 41.37\\
NGC 7469& 0.0163 & S1.5 & 2.70$\pm$0.60\, $\cdot 10^7$          & G13 & 43.60$\pm$0.02 & 1758 & 40.98\\
PG 0844+349& 0.0640 & S1.0 & 7.24$\pm$2.99\, $ \cdot 10^7$          & P04 & ... & 2377 & 41.99\\
\hline
\end{tabular}
\end{center}
\medskip 
Notes: (1) AGN names; (2) redshift from \cite{baumgartner13}; 
(3) source classification from \citet{landt08} or NED
(where S1.0 = Seyfert 1, S1.2-S1.5 = intermediate Seyfert 1, S1n = narrow-line Seyfert 1); 
(4) BH mass (in solar masses) from reverberation mapping campaigns; 
(5) reference for the BH mass, where 
D10: \citet{denney10} but adopting $f=4.31$, G11: \citet{graham11}, G13: \citet{grier13}, P04: \citet{peterson04} but adopting $f=4.31$; 
(6) logarithm of the 14-195 keV intrinsic luminosity \citep[from][]{baumgartner13};
(7) FWHM of the broad emission line component of ${\rm Pa\beta}$
(corrected for instrumental resolution) from \citet{landt08,landt13}.  A 10\% uncertainty has been assumed (see text);
(8) logarithm of the  luminosity of the Pa$\beta$   BLR component from \citet{landt08,landt13}. A 10\% uncertainty has been assumed (see text). 
\noindent
\end{minipage}
\end{table*} 


 \subsection{Fit of the virial relationship using L$_{\rm X}$}\label{sec:fitlx} 
 We have performed a linear fit of the form

%

   \begin{figure}
   \centering
   \includegraphics[width=0.85\hsize]{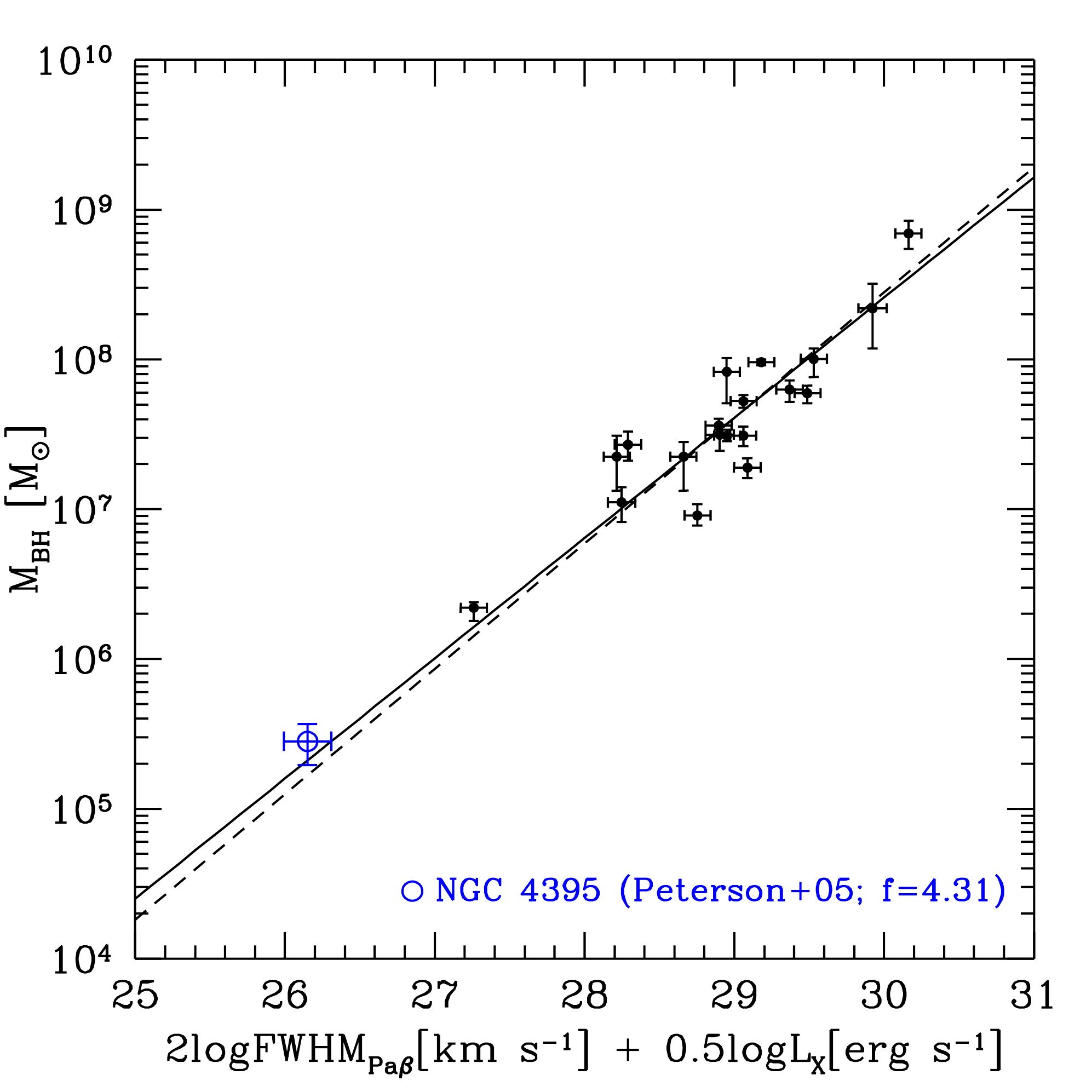}
      \caption{Black Hole masses determined from optical
  reverberation campaigns (adopting a geometrical $f$-factor $f=4.31$) versus the virial product between
  the hard X-ray luminosity ${\rm L_{X}}$ in the 14-195 keV band 
  and the FWHM of the broad component of the Pa$\beta$ emission line. 
  The dashed line shows our best fit using the calibration sample shown in Table \ref{table:1}, while the continuos line is our best fit (eq. \ref{eq:fitlxx}) including also 
  NGC 4395 as measured by \citet{peterson05}, but adopting $f=4.31$  instead of $f=5.5$.
              }
         \label{fig:mlx}
   \end{figure}
%

\begin{equation}
\label{eq:fitlx}
{\rm log}\left(\frac{\rm M_{BH}}{\rm M_{\odot}}\right) =  a \cdot {\rm log}  \left[ \left( {\frac{\rm FWHM_{Pa\beta}}{10^4\, \rm km\,s^{-1}}}\right)^2    \left(   {\frac{\rm L_{\rm X}}{10^{42}\rm\ erg\,s^{-1}}}  \right)^{0.5}     \right]  + b,
\end{equation} 
 where ${\rm M_{BH}}$ is the BH mass, ${\rm FWHM_{Pa\beta}}$ is the measure of the width of  the broad component of the 
 ${\rm Pa\beta}$ emission line, corrected for instrumental broadening, and ${\rm L_{\rm X}}$ is 
 the 14-195 keV intrinsic luminosity. 
 The fit was carried out on a sample of 19 AGN as one galaxy misses the L$_{\rm X}$ measurement (see Table \ref{table:1}).
 We have solved the least-squares 
 problem using the fitting routine FITEXY \citep{press07} 
 that can incorporate errors on both variables and allows us to account for intrinsic scatter. 
 The uncertainties on ${\rm M_{BH}}$ are listed in Table \ref{table:1}, while the errors on the virial product (the x-axis) 
 depend on the uncertainties on the measures of FWHM and ${\rm L_{X}}$. We have calculated the uncertainties on 
 the hard X-ray luminosity ${\rm L_{X}}$ rescaling the 90\% confidence level  
 of the flux in the 14-195 keV band presented in the 70 month Swift/BAT catalogue \citep{baumgartner13}
 to the 68\% confidence level.
 As far as the FWHM measures are concerned, 
 although in some cases the uncertainties are reported in literature, following the studies of 
 \citet{grupe04,vestergaard06,landt08,denney09a}, we have preferred to 
 assume a common uncertainty 
 of 10\%. The best fit provided a slope $a=0.837 \pm 0.040$ and  a y-intercept $b= 7.609 \pm 0.023$ (see dashed line in Figure \ref{fig:mlx}).
 If uncertainties on the FWHM measurements of either 5\% or 20\% are instead assumed, slopes $a=0.814 \pm 0.028$ (where $b=7.617 \pm 0.016$) and $a=0.875 \pm 0.067$ (where $b=7.621 \pm 0.040$)
are obtained, respectively. In the range
$10^5<{\rm M_{BH}/M_{\odot}}<10^{10}$, these two solutions correspond to differences on M$_{\rm BH}$ of $\leq0.10$ dex, if compared to our best fit relation.
 The data have a correlation coefficient $r=0.90$, which corresponds to a probability as low as ${\rm \sim 10^{-7}}$ 
 that they are randomly extracted from an uncorrelated parent population.
 The resulting observed spread is 0.24 dex, while the intrinsic spread (i.e. once the contribution from the
 data uncertainties has been subtracted in quadrature) results to be 0.21 dex. 
 Following \citet{graham13}, 
 we have also used the \citeauthor{tremaine02}'s \citeyearpar{tremaine02} modified version of 
 the routine FITEXY which minimizes the quantity
 \begin{equation}\label{eq:3}
  \chi^2 = \sum_{i=1}^N \frac{(y_i - ax_i -b)}{\sigma_{y_i}^2 + a^2 \sigma_{x_i}^2 + \varepsilon^2} \, ,
 \end{equation}
 where the intrinsic scatter (in the y-direction) is denoted by the term $\varepsilon$, 
 and the measurement errors on the N pairs of observables $x_i$ and $y_i$, 
 (i.e. the virial product ${\rm (2\log FWHM_{Pa\beta} + 0.5 \log L_{X})}$ 
 and the BH masses, respectively) are denoted by $\sigma_{x_i}$ and $\sigma_{y_i}$.
 It results, indeed, that the intrinsic scatter of 0.21 dex makes the reduced $\chi^2$ ($\chi^2/\nu$) from 
 equation \ref{eq:3} (where $\nu$ are the degrees of freedom) equal to $\sim$1. 

We also investigated whether a relationship where M$_{\rm BH}$ depends on L$^{a}$V$^2$ \citep[e.g.][]{vestergaard06} better fits the data.
We obtained $a=0.44 \pm 0.03$, which is similar to the previous fit where the exponent of the dependence on L
is 0.42 (0.837$\times$0.5), however the observed spread, 0.28 dex, resulted to be larger than in the previous fit (0.24 dex)
 and therefore we preferred to use the dependence as shown in eq. \ref{eq:fitlx}. The explanation of this difference resides on the fact
 that  eq. \ref{eq:fitlx} is able to model with a single parameter ($a$) the dependence on both the luminosity and the FWHM of the BLR.

As the AGN are known to vary both in the optical and X-ray bands \citep[e.g.][]{cristiani97,papadakis04,kelly11},
the scatter of the relation should also be affected by the non-simultaneous nature of the X-ray and NIR observations. While the NIR spectra have been taken within  the years 2004 and 2011 \citep{landt08,landt13}, the 14-195 keV X-ray luminosities are the result of the average flux within the years 2005 and 2010 \citep{baumgartner13}. We therefore expect that similar dedicated
projects aimed at observing simultaneously the AGN NIR spectra and their hard X-ray luminosity  have the prospect to obtain even more
accurate relationships to derive the AGN BH mass.

  \subsection{Fit of the virial relationship using L$_{\rm Pa\beta}$} \label{sec:fitlb}

   \begin{figure}
   \centering 
   \includegraphics[width=0.85\hsize]{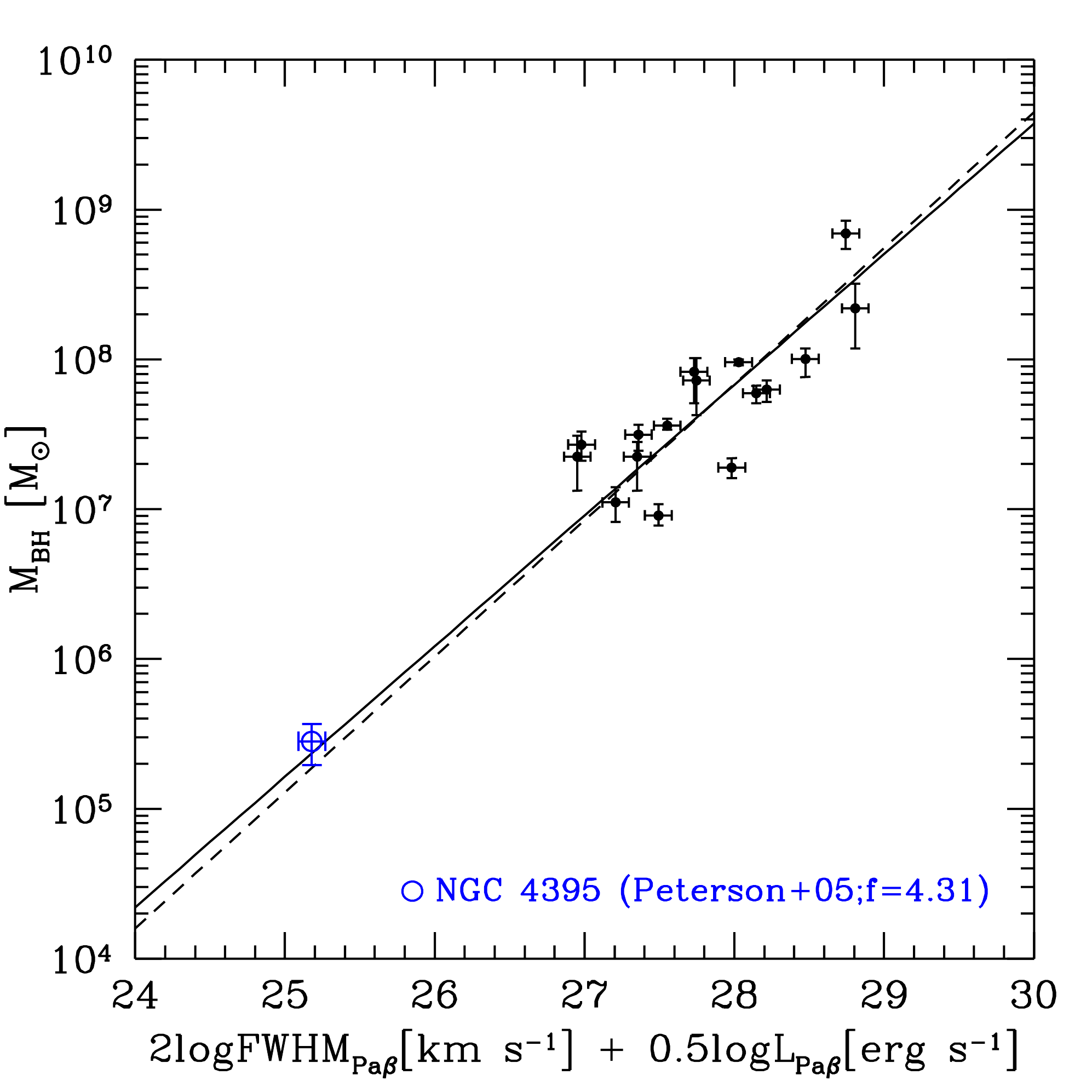}
            \caption{Black Hole masses determined from optical
  reverberation campaigns (adopting a geometrical $f$-factor $f=4.31$) versus the virial product between
  the luminosity ${\rm L_{\rm Pa\beta}}$  and the FWHM of the broad component of the Pa$\beta$ emission line. 
  The dashed line shows our best fit using the calibration sample shown in Table \ref{table:1}, while the continuos line is our best fit (eq. \ref{eq:fitlbb}) including also 
  NGC 4395 as measured by \citet{peterson05}, but adopting $f=4.31$  instead of $f=5.5$.
                       }
         \label{fig:mlb}
   \end{figure}
%

In order to look for a virial relationship able to measure the BH masses using only NIR spectroscopic data,
we have substituted L$_{\rm X}$ with L$_{\rm Pa\beta}$ and then, in analogy with the previous section, have performed a linear fit of the form 

\begin{equation}
\label{eq:fitlb}
{\rm log}\left(\frac{\rm M_{BH}}{\rm M_{\odot}}\right) =  a \cdot {\rm log}  \left[ \left( {\frac{\rm FWHM_{Pa\beta}}{10^4\, \rm km\,s^{-1}}}\right)^2    \left(   {\frac{\rm L_{\rm Pa\beta}}{10^{40}\rm\ erg\,s^{-1}}}  \right)^{0.5}     \right]  + b.
\end{equation} 
In this case the fit was carried out on a sample of 16 AGN as four galaxies miss the L$_{\rm Pa\beta}$ measurement (see Table \ref{table:1}).
As discussed in the previous section, an uncertainty of 10\% on L$_{\rm Pa\beta}$ was used. The best fit provided a slope $a=0.908 \pm 0.060$ and a y-intercept $b= 7.834 \pm 0.031$ (see dashed line in  Figure \ref{fig:mlb}).
 The data have a correlation coefficient $r=0.80$, which has a probability as low as ${\rm \sim 10^{-4}}$ 
 that the data are randomly extracted from an uncorrelated parent population.
 The resulting observed spread is 0.31 dex, while the intrinsic spread results to be 0.28 dex. 
Although this relationship has an intrinsic scatter larger than measured in the previous relation using  L$_{\rm X}$, as already discussed, it has the big advantage that it just needs NIR spectroscopy of the Pa$\beta$ emission line (without using X-ray observations) to derive the AGN BH mass.

  \section{Observations and Line Fit Measurements}

\subsection{NGC 4395}
\label{SectDataNGC}
 We have observed NGC 4395 on December 5 2012, at the LBT with the 
 LBT NIR Spettroscopy Utility with Camera and Integral-Field Unit for Extragalactic Research (LUCI)
 in the zJ band using the grating 200\_H+K in combination with the zJspec filter. A  1$''$${\rm \times}$ 2.8$'$ wide slit was used, 
 corresponding to a ${\rm \sim 220}$ ${\rm km \, s^{-1}}$ spectral resolution. 
 All the observations have been carried out by rotating the slit in order to observe also a bright star that was later used
 to correct for OH absorptions, which are known to vary across the night.
 We acquired 8 images with exposures of 350 s each, using the nodding technique A B B A. The night was clear, 
 the seeing was 0.54$''$ and the airmass was $\sim$1.2. 
 
 Flats and arcs were taken within one day from the observations.
 The data reduction steps included preparation of calibration and science frames, processing and extraction of spectra from science frames, 
 wavelength calibration, telluric correction and flux-calibration. 
 The wavelength calibration made use of the arcs and of a constant correction offset measured on the OH emission lines. 
 The flux-calibration took into account the correction on the telluric flux fraction lost for the seeing. 
 The atmospheric extinction correction has been taken into account 
 automatically during the flux-calibration with the telluric standard star by  using a not extincted stellar model.  
 The spectrum was eventually corrected for Galactic extinction using an extinction value of 0.012 mag in the J band (NED).
 The spectrum of NGC 4395 was extracted with an aperture of 1.5$''$, while the noise spectrum has been derived taking into account both the flux of the scientific object and of the OH airglow.
 
 \subsubsection{The Pa$\beta$ line fit}

\begin{figure}
   \centering
   
   \includegraphics[width=0.85\hsize]{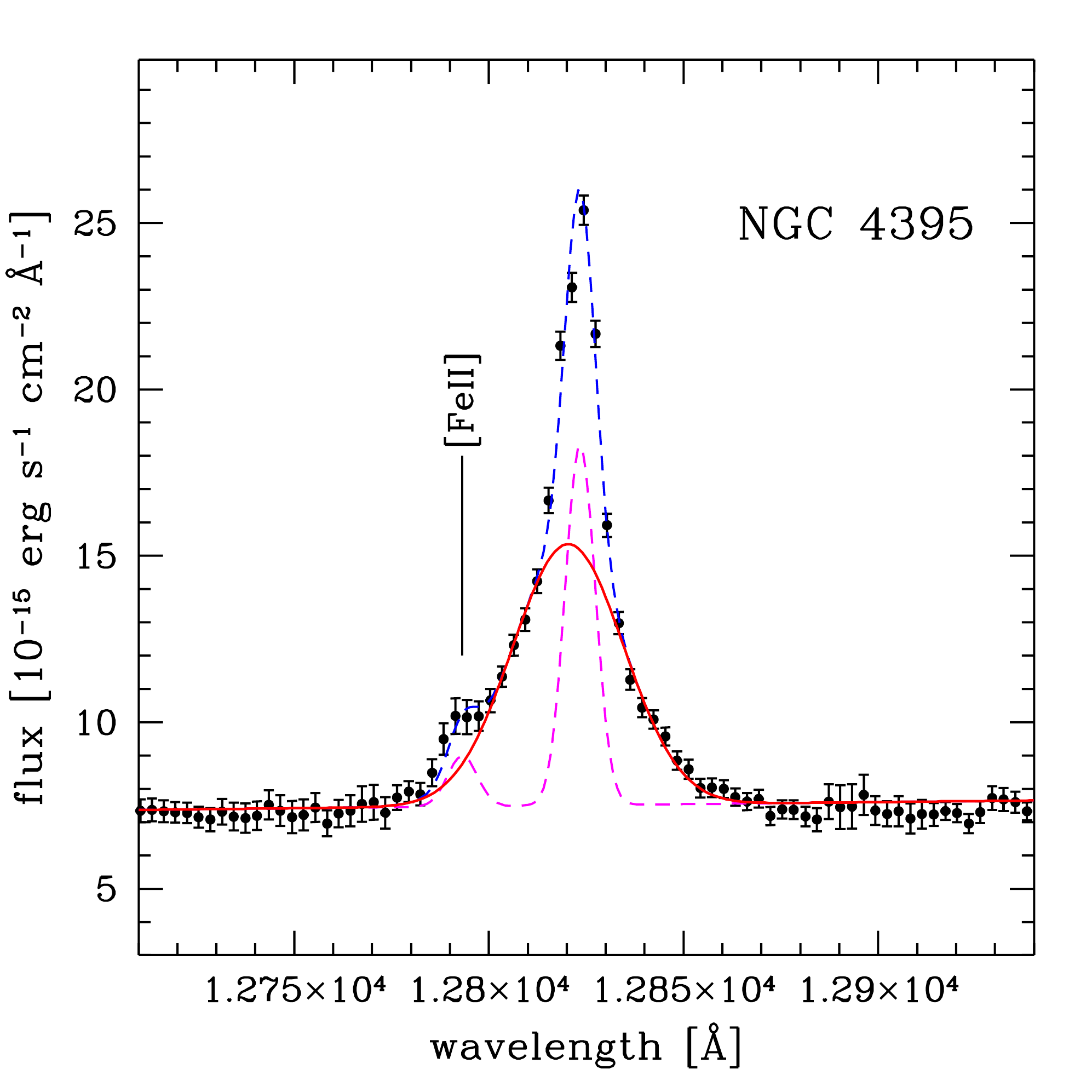}
      \caption{Rest frame, flux calibrated, LUCI@LBT NIR spectrum of NGC 4395  in proximity of Pa$\beta$ emission line. 
 The magenta dashed lines show the narrow 
  component of the Pa$\beta$ and 
  the [FeII] $\lambda$12791
  blended with the Pa$\beta$,  whereas the red solid line is the 
  broad component of Pa$\beta$. 
   The complete fit is shown by a blue dashed line.
              }
         \label{fig:linefit}
   \end{figure}

 We tried to fit the Pa$\beta$ profile by de-blending a broad emission from the narrow one (see Figure \ref{fig:linefit}).
 Although the redshift of NGC 4395 is quite low, the spectrum has been converted to the rest frame.
 The continuum has been estimated with a power law, while the 
 contribution from the Narrow Line Region (NLR) has been fitted with a Gaussian model whose width has been fixed to
 that measured for the forbidden emission line [FeII] $\lambda$12570 (FWHM = 211 km s$^{-1}$). 
 Then, following the NIR transitions listed in \citet{landt08} (see their Table 4), 
 we have added a Gaussian to measure the contamination of the [FeII] $\lambda$12791
 which is in blending with the Pa$\beta$. 
 We have added two Gaussians for the Pa$\beta$: one narrow (as the NLR), and one broad.
 The broad Pa$\beta$ components resulted to have a FWHM of 786$\pm 7$ km s$^{-1}$ and a luminosity of 10$^{38.85\pm 0.02}$ erg s$^{-1}$. After correction for the instrumental broadening it results a FWHM of 755$\pm 7$ km s$^{-1}$.
The fitting uncertainties are quite small ($\sim1$\%), however, as discussed in section \ref{sec:fitlx}, in order to later derive the BH mass, we assumed that the uncertainties on both of these values are 10\%. 
 
 We note that the broad component 
 of the Pa$\beta$ is blueshifted with respect to the narrow one by $\sim3$ $\textrm{\AA}$. Wavelength shifts  of this size, between the BLR and the NLR,
have  already been observed in many other AGNs (Gaskell 1982; Gaskell \& Goosmann 2013). 
 
%

\subsection{MCG -01-24-012}

\begin{figure}
   \centering
   
   \includegraphics[width=0.85\hsize]{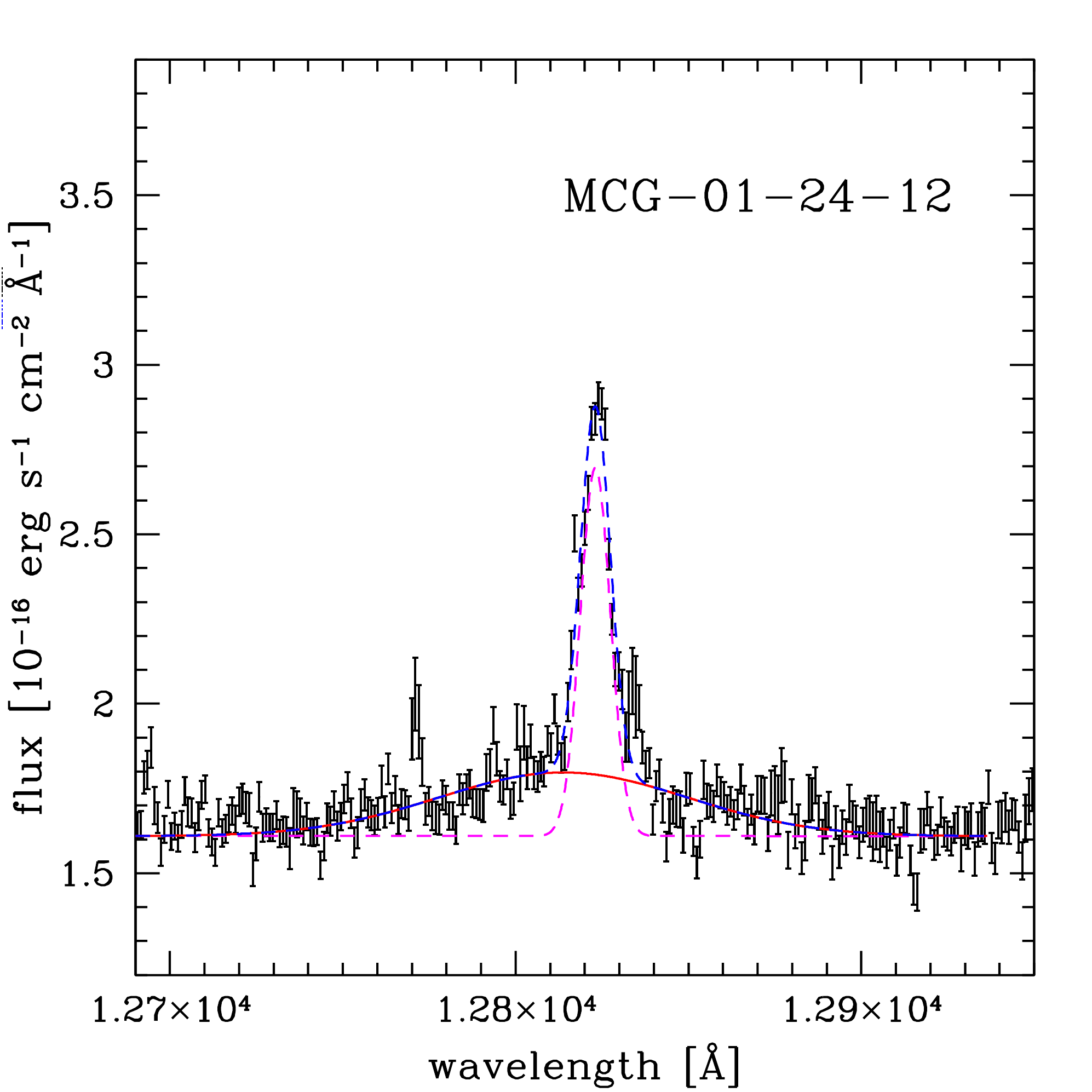}
      \caption{Rest frame, flux calibrated, ISAAC@VLT NIR spectrum of MCG -01-24-012  in proximity of Pa$\beta$ emission line. 
 The magenta dashed lines show the narrow 
  component of the Pa$\beta$ and 
  the [FeII] $\lambda$12791
  blended with the Pa$\beta$,  whereas the red solid line is the 
  broad component of Pa$\beta$. 
   The complete fit is shown by a blue dashed line.
              }
         \label{fig:linefitMCG}
   \end{figure}

We have taken J band (1.1-1.4 $\mu$m) medium resolution (MR) spectra of the nucleus of MCG -01-24-012 on January 7 2012 with ISAAC (Infrared Spectrometer And Array Camera) at the VLT/ESO, in the wavelength range of the Pa$\beta$ line. A 0.8$''$$\times$120$''$  wide slit was used, corresponding to a $\sim60$ km s$^{-1}$ spectral resolution. Four, 340 s long,  spectra were taken. The nodding A B B A technique to remove the sky contribution to the spectra was used.  The night was clear, the seeing was $\sim$1.2$''$ and the airmass  was $\sim$1.6.
We also observed a bright standard star (spectral type: G2V) within 30 minutes from the target observations and with similar airmass, allowing us to use it for the flux calibration and telluric absorption correction.
 As usual, flats and arcs were taken within one day from the observations. Standard data reduction was carried out, similarly as described for NGC 4395 (sect. \ref{SectDataNGC}). The final spectrum was extracted with a spatial aperture of 0.9$''$. The flux calibrated spectrum, in the wavelength region around the Pa$\beta$ emission line, is shown in Figure \ref{fig:linefitMCG}.

%

\subsubsection{The Pa$\beta$ line fit}
\label{SectFitMCG}

The ISAAC  MR spectrum of MCG -01-24-012 shows the [Fe II] $\lambda$12570 and the Pa$\beta$ lines (in Figure \ref{fig:linefitMCG} we show only the region near the Pa$\beta$ line). As usual we converted the wavelength calibration to the rest frame before carrying out the emission line profile fitting. The [Fe II]$\lambda$12570 $\AA$ line profile was modeled using two components of different widths: a narrow one having center at $\lambda_c$=12570.6 \AA\ and  FWHM = 244 km s$^{-1}$, and a broader component with center at $\lambda_c$= 12572.4 and FWHM = 557 km $s^{-1}$, which appears to be redshifted of $\sim$2 \AA\ with respect to the narrow line.  The Pa$\beta$ profile has been modelled with two Gaussian components, a narrow one, with center at $\lambda_c$=12823.1 \AA, having the FWHM tied to the narrow [Fe II] $\lambda$12570 line (in order to take into account of the NLR contribution), and a broad component which, although weak, it is clearly detectable, and results to have the center at $\lambda_c$= 12813.0 \AA, FWHM = 2167$\pm 180$ km s$^{-1}$ (i.e. a 8\% uncertainty) and a flux of 1.74$\pm 0.35$ $\times$10$^{-15}$ erg s$^{-1}$ cm$^{-2}$, corresponding to a luminosity of 10$^{39.18\pm 0.09}$ erg s$^{-1}$.
As discussed in section \ref{sec:fitlx}, and assumed for NGC 4395 (sect. 4.1.1), in order to later derive the BH mass, we used an uncertainty on the FWHM measure of 10\%.   After correcting for the very small instrumental broadening it results a FWHM of 2166 km s$^{-1}$.
We have also tried to add a [Fe II] $\lambda$12791  component in blending with the Pa$\beta$, but its contribution resulted to be negligible. We identified the  2166 km s$^{-1}$ wide component of the Pa$\beta$ line as a  measure of the motion of the BLR gas under the gravitational potential of the central BH. 

 \section{The BH Mass of NGC 4395}

The two above virial relationships can be used to measure the BH mass of NGC 4395. 
Thanks to our NIR observations, it results  
M$_{\rm BH}$=$1.7^{+1.3}_{-0.7} \times 10^5$ M$_{\odot}$  
and
M$_{\rm BH}$=$1.9^{+2.2}_{-1.0} \times 10^5$ M$_{\odot}$    
if the equations based on L$_{\rm X}$ \citep[for NGC 4395 ${\rm L_X = 10^{40.79} ~erg ~s^{-1}}$, with a 1$\sigma$ uncertainty of 9\%;][]{baumgartner13} and L$_{\rm Pa\beta}$ are used, respectively. 
Because of the spreads of the previous fits, as expected, the mass derived using the X-ray luminosity dependent equation is more accurate 
than that measured using the relation dependent from the Pa$\beta$ luminosity. The two estimates are in statistical agreement at less than 1$\sigma$ confidence level, but
are not independent from each other, as they both depend from the same FWHM$_{\rm Pa\beta}$ measurement. 
It is therefore not possible to use both of them to derive a weighted mean estimate of the NGC 4395 BH mass. We then consider the measure 
M$_{\rm BH}$=$1.7^{+1.3}_{-0.7} \times 10^5$ M$_{\odot}$ (based on the more accurate relationship which uses $\sqrt L_{\rm X}$ as a proxy of the BLR radius) as our best estimate of the BH mass in NGC 4395.

As discussed in section \ref{sec:NGC4395}, the BH mass of NGC 4395 has already been estimated by other authors \citep{peterson05,graham11,edri12}.
Our measure is close, and agrees well within the uncertainties, with the above previous measures (once the same geometrical factor $f=4.31$ is adopted). The BH mass measure by \citet{graham11} is based on the same reverberation mapping analysis of the CIV line
by \citet{peterson05}, then, once $f=4.31$ is used, both agree on the value M$_{\rm BH}$=$2.81^{+0.86}_{-0.66} \times 10^5$ M$_{\odot}$ (see 
Figures \ref{fig:mlx} and \ref{fig:mlb},  where our Pa$\beta$ measures and the
 SWIFT/BAT 14-195 keV luminosity have been used to plot the measure by \citet{peterson05}).
The BH mass by 
\citet{edri12} is based on the reverberation mapping analysis of the H$\beta$ line and, once $f=4.31$ is assumed, it corresponds to 
M$_{\rm BH}= (2.69\pm 1.42)\times 10^5$ M$_{\odot}$.  If, instead, our Pa$\beta$ emission-line measures are used together with the relation by \citet[][their eq. 10. See discussion in the next section]{kim10}, a BH mass for NGC 4395 of M$_{\rm BH}$=$1.8^{+1.5}_{-0.8}\times 10^5$ M$_{\odot}$ is obtained. A result very close to our measure. 

 According to \citet{vasudevan07} and \citet{shankar13} the 14-195 keV luminosity ${\rm L_X = 10^{40.79}~erg~s^{-1}}$ of NGC 4395  corresponds to a bolometric luminosity
 ${\rm L_{Bol} = 10^{42.1} ~erg~s^{-1}}$  (assuming $\Gamma=1.8$), and then our BH mass measure implies that the Eddington ratio of NGC 4395 is $\lambda = {\rm log(L_{Bol}/L_{Edd})} = -1.3$. A value quite common in local AGN \citep[see e.g.][]{shankar13}.

\section{Near Infrared SE BH mass relations including NGC 4395}

As the previous independent BH mass estimate of NGC 4395 by \citet{peterson05} is compatible with our measure,
we decided to include the measure by   \citet{peterson05} in the calibration sample and rerun the fits in order to tie and extend the virial SE relationships at masses as low as M$_{\rm BH}\sim 10^5{\rm M}_{\odot}$. This inclusion would not have been possible without our LUCI@LBT spectroscopic observations of the Pa$\beta$ emission line. 
The inclusion of NGC 4395 in the calibration sample does not significantly change the results of the fits. 
If the relationship based on L$_X$ is used,
the best fit provides  (see Figure \ref{fig:mlx})

\begin{eqnarray}
{\rm log (M}_{\rm BH}/{\rm M}_\odot)=0.796(\pm 0.031){\rm log}\left[\left( {\frac{\rm FWHM_{Pa\beta}}{10^4\, \rm km\,s^{-1}}} \right)^2    \left(   {\frac{\rm L_{\rm 14-195 keV}}{10^{42}\rm\ erg\,s^{-1}}}  \right)^{0.5}  \right] \nonumber \\
+ 7.611(\pm0.023)~~ (\pm 0.20)
\label{eq:fitlxx}
\end{eqnarray}
The resulting observed spread is 0.23 dex, while the intrinsic spread results to be 0.20 dex. 
If the relationship based on L$_{\rm Pa\beta}$ is used,
the best fit provides (see Figure \ref{fig:mlb})
\begin{eqnarray}
{\rm log (M}_{\rm BH}/{\rm M}_\odot)=0.872(\pm 0.040){\rm log}\left[\left( {\frac{\rm FWHM_{Pa\beta}}{10^4\, \rm km\,s^{-1}}} \right)^2    \left(   {\frac{\rm L_{\rm Pa\beta}}{10^{40}\rm\ erg\,s^{-1}}}  \right)^{0.5}  \right] \nonumber \\
+ 7.830(\pm0.030)~~ (\pm 0.27)
\label{eq:fitlbb}
\end{eqnarray}
The resulting observed spread is 0.30 dex, while the intrinsic spread results to be 0.27 dex. 
 
 The same kind of relationship has been previously measured by \citet{kim10} using similar data but covering a narrower BH mass range ($10^7 - 10^9$
 M$_\odot$). In order to compare the two relations it is necessary to a) rescale the masses of  \citet{kim10} to a $f=4.31$ geometrical factor instead of the $f=5.5$ used by them and b) correct  the Pa$\beta$ luminosities and FWHM  measured by \citet{landt08,landt13}
 for 1.08 and 0.90 factors (respectively), in order to take into account, following \citet{kim10}, the narrow line component effects in the line profile fits.
 Once these corrections are applied, the two relations result in fairly good agreement within the errors: they give the same results for M$_{\rm BH}\sim 10^{7}$ M$_\odot$, while at M$_{\rm BH}\sim 10^{9}$ M$_\odot$ and M$_{\rm BH}\sim 10^{5}$ M$_\odot$ our estimates are $\sim$0.2 dex smaller and larger, respectively.

 \section{The BH Mass of MCG -01-24-012}   
\label{SectMassMCG}

As already discussed in sect. 2.2, MCG -01-24-012 is an X-ray absorbed Compton thin Seyfert 2 galaxy \citep[N$_H\sim 7 \times 10^{22}$ cm$^{-2}$;][]{malizia02} whose 
intrinsic 14-195 keV X-ray luminosity is ${\rm L_X = 10^{43.55}~ erg~ s^{-1}}$ \citep{baumgartner13}. It should be noted that
in the case of Compton thin
absorption, in the 14-195 keV X-ray band the absorption correction to derive the intrinsic X-ray luminosity is negligible \citep[see, e.g., Figure 11 in][]{burlon11}.
Thanks to the detection of a faint BLR component in the Pa$\beta$ emission line (see sect. \ref{SectFitMCG}) we can calculate the BH mass of the Seyfert 2 galaxy
MCG -01-24-012. If we use eq. \ref{eq:fitlxx}, taking into account a FWHM$_{\rm Pa\beta}$ = 2166 km s$^{-1}$, it results a BH mass of log(M$_{\rm BH}$/M$_\odot$)= 7.17$\pm$0.24 . 
We cannot instead derive the BH mass using eq. 7, as it uses the Pa$\beta$ luminosity which is expected to be severely absorbed in type 2 AGNs.
Indeed, if the observed luminosity of 10$^{39.18}$ erg s$^{-1}$ is used, a mass of log(M$_{\rm BH}$/M$_\odot$)= 6.31$\pm$0.24  is obtained;
an order of magnitude lower than obtained using the X-ray luminosity in eq. 6.

\begin{figure}
   \centering
      \includegraphics[width=0.85\hsize, angle=0]{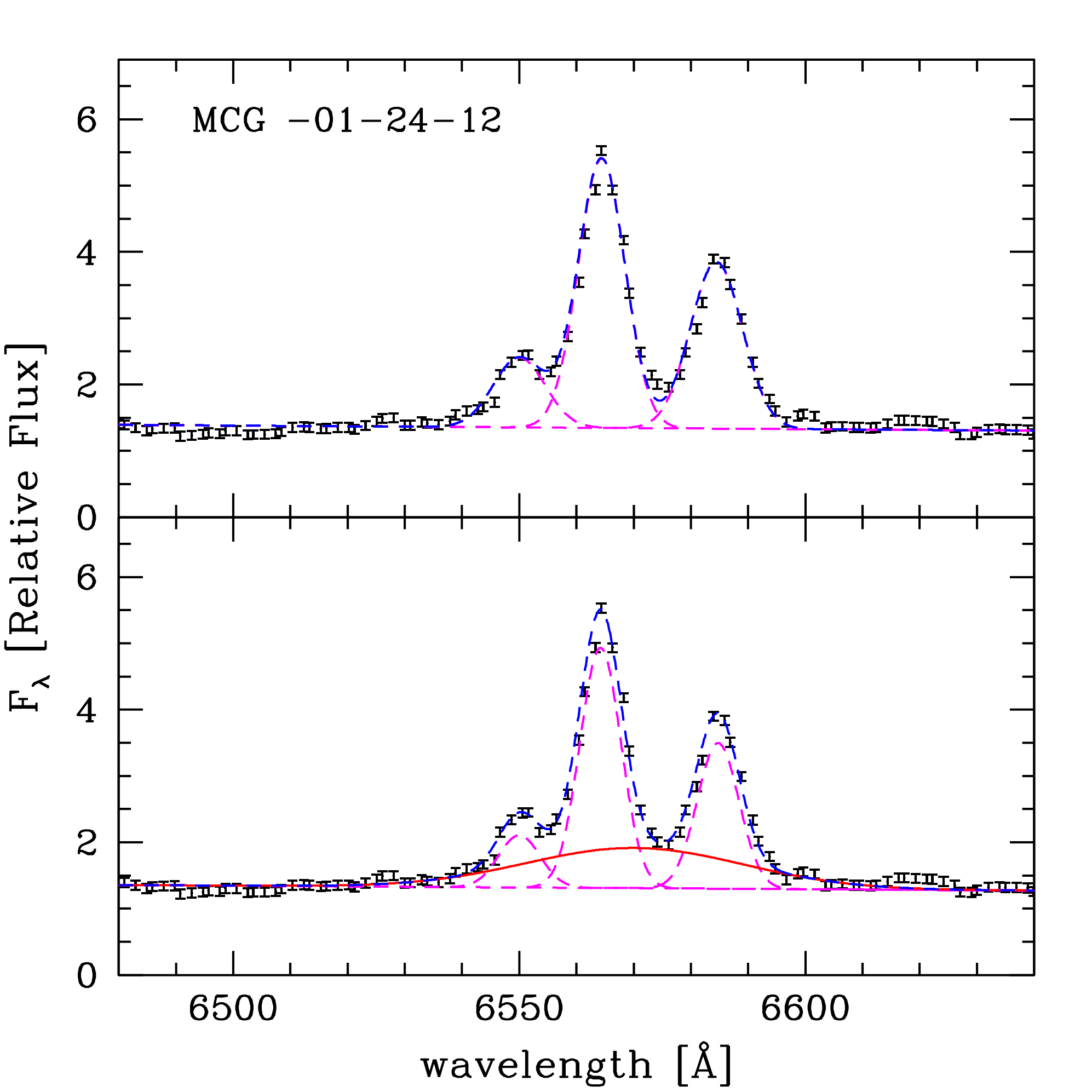}
      \caption{Optical spectrum of MCG -01-24-012 in the wavelength region of the H$\alpha$ emission line. {\it Top.}
      Fit without a broad H$\alpha$ component. {\it Bottom.} Fit including also a broad (FWHM $\sim$ 20$\times$10$^2$ km s$^{-1}$) H$\alpha$ component. 
              }
         \label{fig:linefitMCGHa}
   \end{figure}

Although MCG -01-24-012 is classified as a Seyfert 2 galaxy we checked whether a faint broad H$\alpha$ component was detectable in  the 6dF optical spectrum \citep[having resolution $R=\lambda/\Delta\lambda\sim$550;][and references therein]{jones09}. 
The H$\alpha$ and [NII] emission lines  are roughly well reproduced using narrow components only, all having widths compatible with the instrumental spectral resolution (FWHM $\sim$ 550 km s$^{-1}$).
However, for the sake of completeness, it should be noted that a significantly better fit is obtained if  a broad H$\alpha$ component, having FWHM = $2.0\pm1.8\times 10^3$ km s$^{-1}$, is included (see Figure \ref{fig:linefitMCGHa}). The optical data have not a very good  spectral resolution and therefore the FWHM estimate is quite uncertain. A proper subtraction of a starlight component could be very important  in testing
for the presence of a weak broad H$\alpha$ component. However the quality of the data are not sufficient to carry out a starlight subtraction.
If we use the measure
of the width of the broad component of the H$\alpha$ line and convert the 14-195 keV X-ray luminosity into a 2-10 keV luminosity, we can use eq. 5 in \citet{bongiorno14} to estimate the BH mass. If a photon index $\Gamma$=1.8 is assumed, log(M$_{\rm BH}$/M$_\odot$)= $7.1^{+0.6}_{-2.1}$  is obtained
(a change of $\Delta\Gamma$= $\pm$0.1 in the assumed photon index corresponds to a variation of $\rm \Delta logM_{\rm BH}$/M$_\odot$= $\pm$0.08 in the BH mass).
The large uncertainty on the BH mass measure based on the H$\alpha$ line prevents a  significant comparison with the estimate based on the Pa$\beta$ line. Nonetheless, it should be noted that the two best fit values are very similar. 

As done for NGC 4395, the 14-195 keV luminosity ${\rm L_X = 10^{43.55} ~ erg~ s^{-1}} $ of MCG -01-24-012  can be converted to a bolometric luminosity L$_{\rm Bol}$= 10$^{44.8}$ erg s$^{-1}$. Therefore our BH mass measure implies that the Eddington ratio of MCG -01-24-012 is $\rm \lambda = {\rm log(L_{Bol}/L_{Edd})} = -0.5$. A value which is, also in this case, quite typical
for local AGN \citep{shankar13}.

%

\section{Luminosity of the Pa$\beta$ as a function of the 14-195 keV Luminosity}   

\begin{figure}
   \centering
   \includegraphics[width=0.85\hsize]{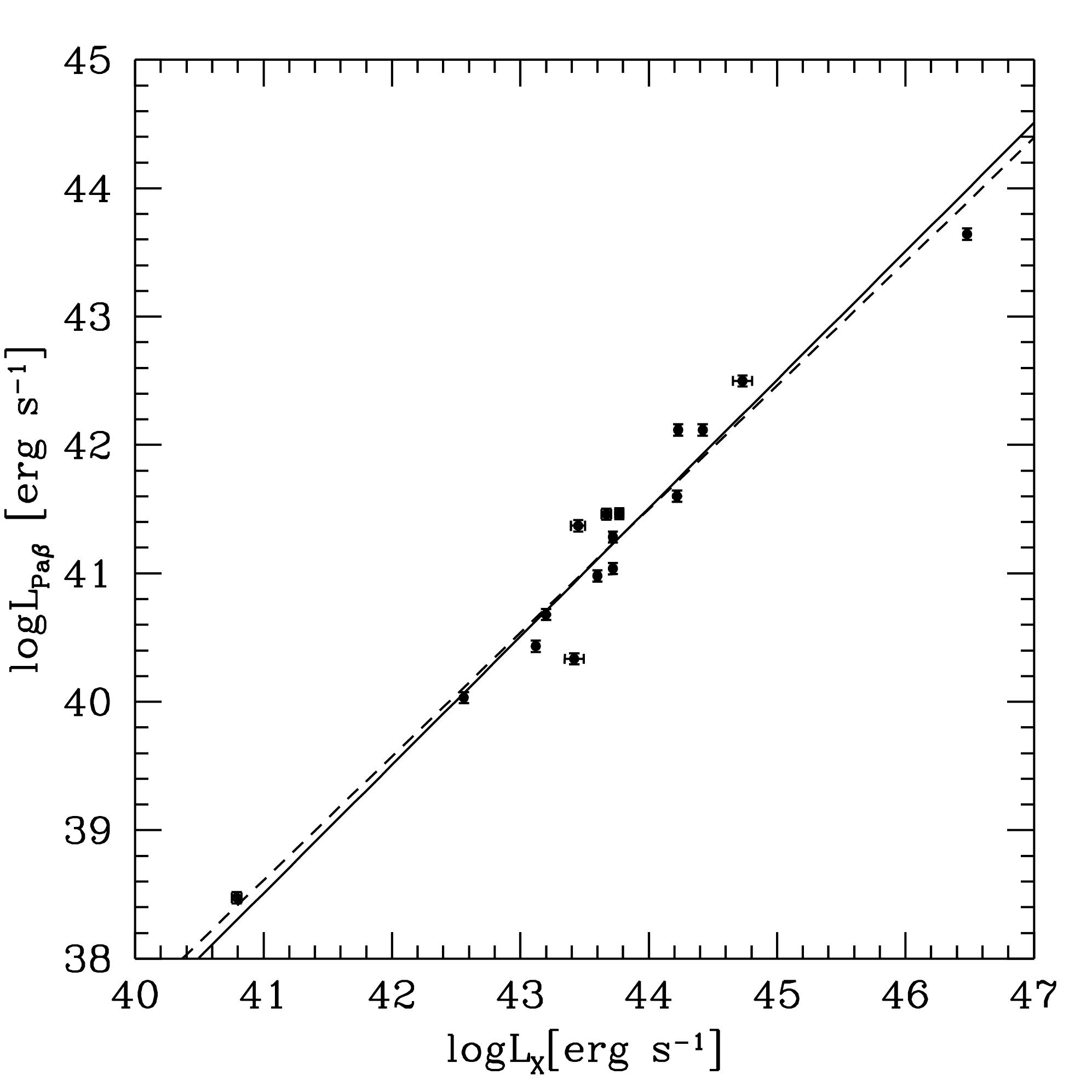}
      \caption{Luminosity of the BLR component of the Pa$\beta$ as a function of the intrinsic
       14-195 keV band luminosity, L$_{\rm X}$. The dashed line is our best fit with free slope, while the black continuous line is
       our best fit with unitary slope (eq. \ref{eq:lblx}).
                    }
         \label{fig:lblx}
   \end{figure}

The two above discussed virial relationships (eqs. \ref{eq:fitlxx} and \ref{eq:fitlbb}) imply that a significant correlation should exist between the Pa$\beta$ emission line luminosity (the BLR component), L$_{Pa\beta}$, and the 14-195 keV luminosity, L$_X$.
Similar kind of relationships have been already studied and derived by other authors. For example, \citet{panessa06} studied the correlation between the H$\alpha$ luminosity and the 2-10 keV luminosity on a sample of 87 Seyfert galaxies and low-z QSOs, finding 
\begin{equation}
{\rm logL_{2-10 keV} = 1.06 (\pm 0.04) \cdot logL_{H\alpha}  -1.14(\pm 1.78)} .
\label{eq:panessa}
\end{equation}
Using our calibration sample (with NGC 4395 included) we performed a free slope fit of the relation 
\begin{equation}
{\rm log\left({L_{Pa\beta}\over{10^{42} erg~s^{-1}}} \right)= a\cdot log\left({L_{14-195 keV}\over{10^{44}erg~s^{-1}} }\right) + b.} 
\label{eq:lblx}
\end{equation}
The  data ($\rm logL_{Pa\beta}$ and $\rm logL_{\rm X}$) have a correlation coefficient $r= 0.97$ which corresponds to a  probability of $\sim 10^{-10}$ of being drawn from an un-correlated population, while the fit provides a slope $a=0.963\pm 0.011$ and a y-intercept $b=-0.500\pm 0.013$ (dashed line in Figure \ref{fig:lblx}), with an intrinsic spread of 0.26 dex. If a unitary slope is assumed, then our fit provides:

\begin{equation}
{\rm log\left({L_{Pa\beta}\over{10^{42} erg~s^{-1}}} \right)= log\left({L_{14-195 keV}\over{10^{44}erg~s^{-1}} }\right) -0.492(\pm 0.013) \pm  0.26} 
\label{eq:lblx}
\end{equation}
(see the black continuous line in Figure \ref{fig:lblx}). If
we assume a photon index $\Gamma = 1.8$ and a ratio of  $\rm log( {Pa\beta}/{H\alpha})= -1.2$ (case B recombination),
our fit is in partial agreement with eq. \ref{eq:panessa}  \citep[from][]{panessa06}.
Their eq.    \ref{eq:panessa} should read  ${\rm logL_{Pa\beta} = 0.943\cdot logL_{14-195 keV}  -0.5}$, which predicts, in the range ${\rm 42\leq logL_{14-195 keV} \leq 46}$, BLR Pa$\beta$ luminosities $\sim$0.4 dex lower than on average observed in our data. This discrepancy could be attributed to the assumption of a case B recombination ratio
between the Pa$\beta$ and H$\alpha$ lines. There is indeed evidence that, in AGN, the Paschen line ratios rule out case B recombination \citep{soifer04,glikman06}. 

\section{Conclusions}   

As discussed in the introduction, the measure of local, low mass, BHs,  such as those in the Seyfert 1 galaxy NGC  4395 and the Seyfert 2 galaxy MCG -01-24-012, is of great importance as 
these BHs give important informations on the AGN/galaxy formation and evolution.  NGC 4395 hosts
one of the local, smallest, BH of AGN origin ever found. Previous measures (if $f$=4.31 is used) obtained M$_{\rm BH}$=$2.81^{+0.86}_{-0.66} \times 10^5$ ${\rm M}_{\odot}$   from a reverberation mapping analysis of the CIV emission line \citep{peterson05} and $\rm M_{\rm BH} = (2.69\pm 1.42)\times 10^5$ ${\rm M}_{\odot}$ from a reverberation mapping analysis of the H$\beta$ emission line \citep{edri12}.

We have derived two new SE BH mass relationships based on the FWHM and luminosity of the BLR component of the Pa$\beta$ emission line and on the
hard X-ray luminosity in the 14-195 keV band. These relationships have been initially calibrated in the $10^7 - 10^9$ M$_\odot$ mass range and have then been used to obtain a new independent BH mass measure of NGC 4395. We obtained  M$_{\rm BH}$=1.7 $^{+1.3}_{-0.7} \times 10^5$ $  {\rm M}_{\odot}$, which resulted in agreement ($<$1 $\sigma$) with the two previous existing estimates,
therefore confirming the very small mass of its BH.

Thanks to our new measure of the Pa$\beta$ line and then mass of NGC 4395 we were able to extend in the $10^5 - 10^9$ M$_\odot$ mass range our SE relationships. These relations are able to reliably measure the BH mass on those obscured or low luminosity AGN where the nuclear component is less visible and/or contaminated by the hosting galaxy. 
Our method has been indeed also successfully applied to the Seyfert 2 galaxy MCG -01-24-012, whose BH mass has been measured for the first time
to be M$_{\rm BH}$=1.5 $^{+1.1}_{-0.6} \times 10^7$ $  {\rm M}_{\odot}$.

These results were already implicitly suggested by the study of \citet{lafranca14}  where it is discussed how the observed  correlation between the
AGN X-ray variability and the BH mass \citep[e.g.][]{ponti12,kelly13} is equivalent to a correlation between the X-ray variability and the virial product ($\sqrt L V^2$) computed using the same Pa$\beta$ and the hard X-ray luminosity measures utilised in this work.

We can conclude that our new derived NIR and hard X-ray based SE relationships could be
of great help in measuring the BH mass in low luminosity and absorbed AGN and therefore better measuring the complete (AGN1+AGN2) super massive BH mass function. In this respect, in the future, a similar technique could also be applied at larger redshift. For example, at redshift $\sim$2-3  the Pa$\beta$ line could be observed in the 1-5 $\mu$m wavelength range with NIRSPEC on the James Webb Space Telescope\footnote{See http://www.stsci.edu/jwst/instruments/nirspec}. While, after a recalibration, the rest frame 14-195 keV X-ray luminosity could substituted by  the 10-40 keV hard X-ray band (which is as well not so much affected by obscuration for mildly absorbed, Compton-thin, AGN). At redshift $\sim$2-3,
in the observed frame, the 10-40 keV hard band roughly corresponds to the 2-10 keV energy range which is typically observed with the Chandra and
XMM-Newton telescopes.

\section*{Acknowledgments}
We thank the referee for  her/his very useful comments which improved the manuscript.
Based on observations made with ESO telescopes at the Paranal Observatory and the Large Binocular Telescope (LBT) at Mt. Graham, Arizona. The LBT  is an international collaboration among institutions in the United States, Italy, and Germany. LBT Corporation partners are: The University of Arizona on behalf of the Arizona university system; Istituto Nazionale di Astrofisica, Italy; LBT Beteiligungsgesellschaft, Germany, representing the Max-Planck Society, the Astrophysical Institute Potsdam, and Heidelberg University; The Ohio State University; and The Research Corporation, on behalf of The University of Notre Dame, University of Minnesota, and University of Virginia.

We thank for the very efficient service observing mode operation management at LBT and the ESO staff at Paranal and Garching for their support during the preparation and execution of the observations. MB acknowledges support from the FP7 Career Integration Grant "eEASy" (CIG 321913).
We acknowledge funding from PRIN-MIUR 2010 award 2010NHBSBE and from PRIN-INAF 2011 ``BH growth and AGN feedback through the cosmic time''.

\bibliographystyle{mn2e} 
\bibliography{mybib} 

\label{lastpage}

\end{document}